\documentclass[12pt,preprint]{aastex}
\usepackage{epsfig}
\usepackage{amsmath}
\usepackage{psfig}
\def\msun{M_{\odot}}
\shorttitle{{Fundamental Properties of Gamma-rays from Globular Clusters}}
\shortauthors{Hui et al.}
\begin{document}

\title{The Fundamental Plane of Gamma-ray Globular Clusters}

\author{
C. Y. Hui\altaffilmark{1}, K. S. Cheng\altaffilmark{2}, Y. Wang\altaffilmark{2},
P. H. T. Tam\altaffilmark{3}, A. K. H. Kong\altaffilmark{3},
D. O. Chernyshov\altaffilmark{4} and V. A. Dogiel\altaffilmark{5}
}

\altaffiltext{1} {Department of Astronomy and Space Science, Chungnam National University,
Daejeon, South Korea}
\altaffiltext{2}
{Department of Physics, University of Hong Kong, Pokfulam Road, Hong
Kong}
\altaffiltext{3}
{Institute of Astronomy and Department of Physics, National Tsing Hua University, Hsinchu, Taiwan}
\altaffiltext{4}
{Moscow Institute of Physics and Technology, Institutskii lane, 141700
Moscow Region, Dolgoprudnii, Russia.}
\altaffiltext{5}
{I.E.Tamm Theoretical Physics Division of P.N.Lebedev
Institute, Leninskii pr, 53, 119991 Moscow, Russia}

\begin{abstract}
We have investigated the properties of a group of $\gamma$-ray emitting globular
clusters (GCs) which have recently been uncovered in our Galaxy. By correlating the
observed $\gamma$-ray luminosities $L_{\gamma}$ with various cluster properties, we probe the
origin of the high energy photons from these GCs. We report $L_{\gamma}$
is positively correlated with the encounter rate $\Gamma_{c}$ and the metalicity
$\left[{\rm Fe/H}\right]$ which place
an intimate link between the gamma-ray emission and the millisecond pulsar
population. We also find a tendency that $L_{\gamma}$
increase with the energy densities of the soft photon at the cluster location.  
Furthermore, the two-dimensional regression analysis suggests that $L_{\gamma}$, soft photon densities, 
and $\Gamma_{c}$/$\left[{\rm Fe/H}\right]$ possibly span fundamental planes which 
potentially provide better predictions for the $\gamma$-ray properties of GCs.
\end{abstract}


\section{INTRODUCTION}
Millisecond pulsars (MSPs) are generally believed as the descenders of the low-mass
X-ray binaries (LMXBs) (Alpar et al. 1982). As the formation rate per unit mass of
LMXBs is orders of magnitude greater in globular clusters (GCs) than in the Galactic
field (Katz 1975; Clark 1975), it is not surprise that $80\%$ of the detected MSPs
are located in GCs\footnote{see also http://www.naic.edu/$\sim$pfreire/GCpsr.html} 
(cf. Manchester et al. 2005). The relatively high formation rate of LMXBs and MSPs is a
natural consequence of the frequent stellar encounters. With the X-ray populations
in various GCs have been revealed by the \emph{Chandra X-Ray Observatory}, Pooley
et al. (2003) and Gendre et al. (2003) have found a positive correlation between
the number of LMXBs in GCs and the stellar encounter rate, $\Gamma_{c}$. This
provides evidence for the dynamical formation of LMXBs in GCs. As the descenders
of the LMXBs, MSPs are also expected to have a dynamically origin.

Very recently, with the corrections of the observational effects in the radio pulsar
surveys toward different GCs, Hui, Cheng \& Taam (2010) have found a positive correlation
between the MSP populations in GCs and $\Gamma_{c}$, which has long been predicted. 
Moreover, the authors have also found another positive correlation between the
metalicity and the MSP population. This relation is not unexpected as the
high metalicity in a GC can
result in a more efficient orbital shrinkage by magnetic braking. Therefore, the
parameter space for the successful Roche-lobe overflow is enhanced (Ivanova 2006)
and subsequently lead to a higher formation rate of MSPs.

A brand new window for investigating the MSPs in GCs has been open by the launch of
the \emph{Fermi Gamma-ray Space Telescope}. Since MSPs are the only steady $\gamma$-ray 
emitters in GCs, observations with \emph{Fermi} can provide an alternative channel for 
investigating MSP populations. Shortly after the commence of its
operation, the Large Area Telescope (LAT) onboard the spacecraft has detected
$\gamma-$rays from 47 Tucanae (hereafter 47~Tuc) (Abdo et al. 2009).
Terzan~5, which hosts the largest
known MSP population, has also been subsequently detected (Kong et al. 2010).
As the sensitivity of LAT increases monotonically with the continuous all-sky
survey, a total of 15 confirmed detections of $\gamma-$ray emitting GCs have very recently been reported by
Abdo et al. (2010a; 2010b) and Tam et al. (2010). Using $\sim1.5$ years of LAT data, Abdo et al. (2010a) have 
detected 6 new $\gamma$-ray GCs besides 47~Tuc and Terzan~5. On the other hand, 
Tam et al. (2010) have recently reported 7 other new detections with $\sim2$ years data. Among all known 
$\gamma$-ray GCs, Liller~1, which has the highest metalicity in our Milky way, is also found to 
have the highest $\gamma$-ray luminosity (Tam et al. 2010). This discovery further suggests that the effects of 
metalicity cannot be neglected. Thanks to these surveys, we are able to 
study these clusters as a unique class for the first time.

To explain the $\gamma-$rays from GCs, there are two main streams. Venter \& de Jager
(2008) and Venter et al. (2009) suggest the $\gamma-$rays are originated from the
curvature radiation of electrons in MSP magnetospheres. However, the frequent stellar
interactions can lead to a complicated magnetic field structure (Cheng \& Taam 2003),
which can possibly explain the difference of the radio and X-ray properties of MSPs in
GCs with respect to those located in the Galactic field (cf. Hui, Cheng
\& Taam 2009, 2010). One consequence of the complicated surface magnetic field is to 
turn off the accelerating region for producing high energy photons. Ruderman \& Cheng (1988) 
argue that if the open field lines are curving upward due to the effect of local field 
then in this case $e^{-}/e^{+}$ pair production and outflow can occur on all open-field 
lines. As a result, the outer-magnetospheric gap is quenched by these pairs. This scenario 
is supported by the fact that the MSPs in 47~Tuc are essentially thermal X-ray emitters 
(Bogdanov et al. 2006). All these demonstrate the potential difficulties of the pulsar 
magnetospheric model in explaining the observed $\gamma-$rays from GCs, which motivate 
the exploration of additional / alternative emission mechanisms. 

On the other hand, Bednarek \& Sitarek (2007) have proposed
that the inverse Compton scattering (ICS) could be a possible mechanism to produce
$\gamma-$rays from GCs. In their
model they predict that GCs could be sources of GeV-TeV photons. Unfortunately
they have ignored the contribution from the Galactic background photons.
Very recently, through generalizing the ICS model by including various soft photon
fields, Cheng et al. (2010) have found that the observed $\gamma-$ray
spectra of all 8 $\gamma$-ray GCs reported by Abdo et al. (2010a) can be well-modeled  
by the ICS between relativistic electrons/positrons in the pulsar wind of MSPs
in the GCs and the background Galactic soft photons. This provides another possible
explanation for the origin of the $\gamma-$rays.

In this paper, we report the results from exploring the $\gamma-$ray emission
properties by comparing with various cluster properties, which provides us with insight on the
origin of the $\gamma-$rays from this class of GCs. In \S2, we report the method and
the results from the correlation and the regression analysis. We subsequently discuss the implication
of these results in \S3.

\section{CORRELATION \& REGRESSION ANALYSIS}

Abdo et al. (2010a) and Tam et al. (2010) have reported 15 GCs with firm detections. This sample size is 
somewhat larger than that adopted by Pooley et al. (2003) for investigating the relations between the 
X-ray point source populations in GCs and various cluster parameters. 
The sensitivity limit of the current sample is $\sim6\times10^{-12}$~erg~cm$^{-2}$~s$^{-1}$ (0.1-100~GeV).
The properties of 15 confirmed $\gamma-$ray GCs are summarized in Table~\ref{gc_info} and the
entries are explained in the following. 

\begin{deluxetable}{lccccccc}
\tablewidth{0pc}
\tablecaption{Properties of the $\gamma-$ray emitting GCs.}
\startdata
\hline\hline
Cluster Name & $d$\tablenotemark{a} & $\Gamma_{\rm c}$\tablenotemark{b} & [Fe/H]\tablenotemark{c} &
$M_V$\tablenotemark{d} &
$u_{\rm optical}$\tablenotemark{e} & $u_{\rm IR}$\tablenotemark{e} 
& $\log L_{\gamma}$\tablenotemark{f} \\
{}  & kpc &  &  & & eV~cm$^{-3}$ & eV~cm$^{-3}$ & erg~s$^{-1}$ \\\hline\hline
\multicolumn{8}{c}{Adopted from Abdo et al. (2010a)}\\\hline
47 Tuc  & 4.0 & 44.13 & -0.76 & -9.17 & 0.93 & 0.25 & $34.68^{+0.12}_{-0.13}$\\
Omega Cen & 4.8 & 4.03 & -1.62 & -10.07 & 1.61 & 0.51 & $34.44^{+0.13}_{-0.15}$\\
M 62  & 6.6 & 47.15 & -1.29 & -9.09 & 8.07 & 0.86 &  $35.04^{+0.12}_{-0.14}$\\
NGC 6388  & 11.6 & 101.99 & -0.60 & -9.74 & 2.59 & 0.56 & $35.41^{+0.12}_{-0.25}$\\
Terzan 5  & 5.5 & 118.29 & 0.00 & -6.51 & 7.02 & 1.37 & $35.41^{+0.17}_{-0.19}$\\
NGC 6440  & 8.5  & 74.17 & -0.34 & -8.78 & 10.79 & 1.00 & $35.38^{+0.19}_{-0.15}$\\
M 28  & 5.1 & 13.10 & -1.45 & -7.98 & 5.47 & 0.92 & $34.79^{+0.16}_{-0.16}$\\
NGC 6652  & 9.0 & 1.24 & -0.96 & -6.43 & 3.65 & 0.51 & $34.89^{+0.18}_{-0.15}$\\\hline
\multicolumn{8}{c}{Adopted from Tam et al. (2010)}\\\hline
Liller 1  & 9.6 & 77.98  & 0.22 & -7.63 & 10.53 & 1.40 & $35.77^{+0.13}_{-0.18}$  \\
M 80      & 10.3 & 31.31 & -1.75 & -8.23 & 1.88 & 0.33 & $34.92^{+0.28}_{-0.51}$  \\
NGC 6441  & 11.7 & 88.42 & -0.53 & -9.64 & 3.59 & 0.69 & $35.57^{+0.09}_{-0.12}$  \\
NGC 6624  & 7.9 & 14.65 & -0.44 & -7.49 & 6.03 & 0.67 &  $35.17^{+0.09}_{-0.11}$ \\
NGC 6541  & 6.9 & 20.00 & -1.83 & -8.34 & 4.69 & 0.61 &  $34.54^{+0.24}_{-0.33}$ \\
NGC 6752  & 4.4 & 10.78 & -1.56 & -7.94 & 2.01 & 0.48 &  $34.14^{+0.19}_{-0.30}$ \\
NGC 6139  & 10.1 & 13.28 & -1.68 & -8.36 & 4.10 & 0.69 & $35.03^{+0.19}_{-0.34}$  
\enddata
\tablenotetext{a}{\footnotesize Cluster distance adopted from Abdo et al. (2010a) and Tam et al. (2010).}
\tablenotetext{b}{\footnotesize Two-body encounter rate estimated by $\rho_{0}^{2}r_{c}^{3}\sigma_{0}^{-1}$ with
the value scaled with that in M4 which has $\rho_{0}=10^{3.82}$~$L_{\odot}$pc$^{-3}$,
$r_{c}=0.53$~pc and $\sigma_{0}=8.9$~km/s}.
\tablenotetext{c}{\footnotesize Metalicity}
\tablenotetext{d}{\footnotesize Absolute visual magnitude}
\tablenotetext{e}{\footnotesize Energy densities of various soft photon fields (see text)}
\tablenotetext{f}{\footnotesize $\gamma-$ray luminosities adopted from in Abdo et al. (2010a) and 
Tam et al. (2010)}
\label{gc_info}
\end{deluxetable}

For choosing the cluster parameters for the correlation analysis, we have considered two-body encounter
rate $\Gamma_{c}$, metalicity $\left[{\rm Fe/H}\right]$, absolute visual magnitude $M_{V}$, as well as 
Galactic background optical / infrared photon densities at the locations of the GCs $u_{\rm optical}$ / $u_{\rm IR}$.

\begin{figure*}
\centering
\psfig{figure=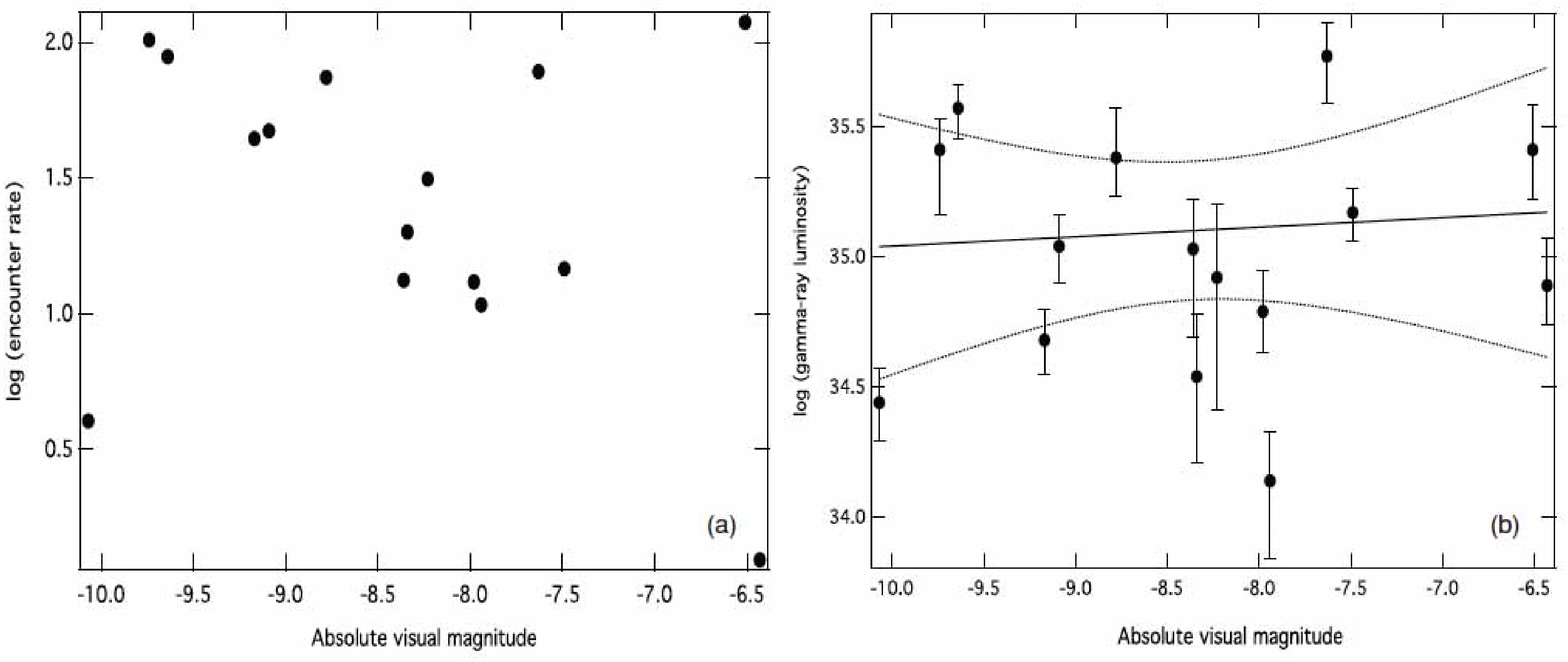,width=18cm,clip=}
\caption[]{{\bf a.} Two-body encounter rate $\Gamma_{c}$ vs. absolute visual magnitude $M_{V}$. {\bf b.} The $\gamma-$ray luminosity 
$L_{\gamma}$ vs. $M_{V}$. The straight line represents the best-fit straight line 
with the errors of the data points fully taken into account. The dotted lines represent
the upper and the lower $95\%$ confidence bands.} 
\label{mass}
\end{figure*}

$\Gamma_{\rm c}$ is the most obvious parameter related to the binary formation rate and hence the
number of MSP in a GC. This parameter
can be estimated as $\rho_{0}^{2}r_{\rm c}^{3}\sigma_{0}^{-1}$ where $\rho_{0}$ is the central luminosity density,
$r_{\rm c}$ is the core radius and $\sigma_{0}$ is the velocity dispersion at the cluster center.
$\sigma_{0}$ are adopted from Gnedin et al. (2002). For $\rho_{0}$ and $r_{\rm c}$, the values are taken from
Harris (1996; 2003 version) and modified for the distances adopted for this analysis (cf. Tab.~\ref{gc_info}).
Besides $\Gamma_{\rm c}$, Hui et al. (2010) have shown that $\left[{\rm Fe/H}\right]$ are also
a key parameter in determining the size of the MSP population in a GC. The values of
$\left[{\rm Fe/H}\right]$ are taken from Harris (1996). 
On the other hand, if stellar encounters were not the major channel of the binary formation, one would
expect the binary population to be correlated with the cluster mass $M_{\rm GC}$ for a primordial binary
origin (Lu et al. 2009; Lan et al. 2010). Pooley et al. (2003) have estimated the cluster mass by 
integrating the King's profiles of the GCs. And therefore, their mass estimates are naturally correlated with 
$\Gamma_{c}$. On the other hand, 
assuming a constant mass-to-light ratio, $M_{\rm GC}$ can also be estimated from the absolute visual
magnitude $M_{V}$: $M_{\rm GC}=10^{-0.4M_{V}}$ (cf. Hui et al. 2010; Lu et al. 2009). Different from the estimates adopted 
by Pooley et al. (2003), the correlations between our mass estimates with $\Gamma_{c}$ is only confident at the level 
less than $53\%$ (see Figure~\ref{mass}a). Also, the correlation between $M_{V}$ and [Fe/H] only attains a confidence level 
$\lesssim56\%$. 
The values of $M_{V}$ are also taken from Harris (1996) and modified for the adopted distances as presented in 
Table~\ref{gc_info}.

Apart from the number of MSPs, in the context of ICS model, the $\gamma-$ray luminosity of a GC also depends on the
energy density of the soft photon field (see Cheng et al. 2010). There are three components of
background photons in the Galaxy which can interact with the relativistic leptons:
they are relic, infrared and optical
photons. As the energy density of the relic photons is uniform and does not vary from cluster
to cluster, we ignore it in our analysis. We obtain the estimates of Galactic optical and infrared photon
density, $u_{\rm optical}$ and $u_{\rm IR}$ with the GALPROP code (Strong \& Moskalenko 1998). 

\begin{figure*}
\centering
\psfig{figure=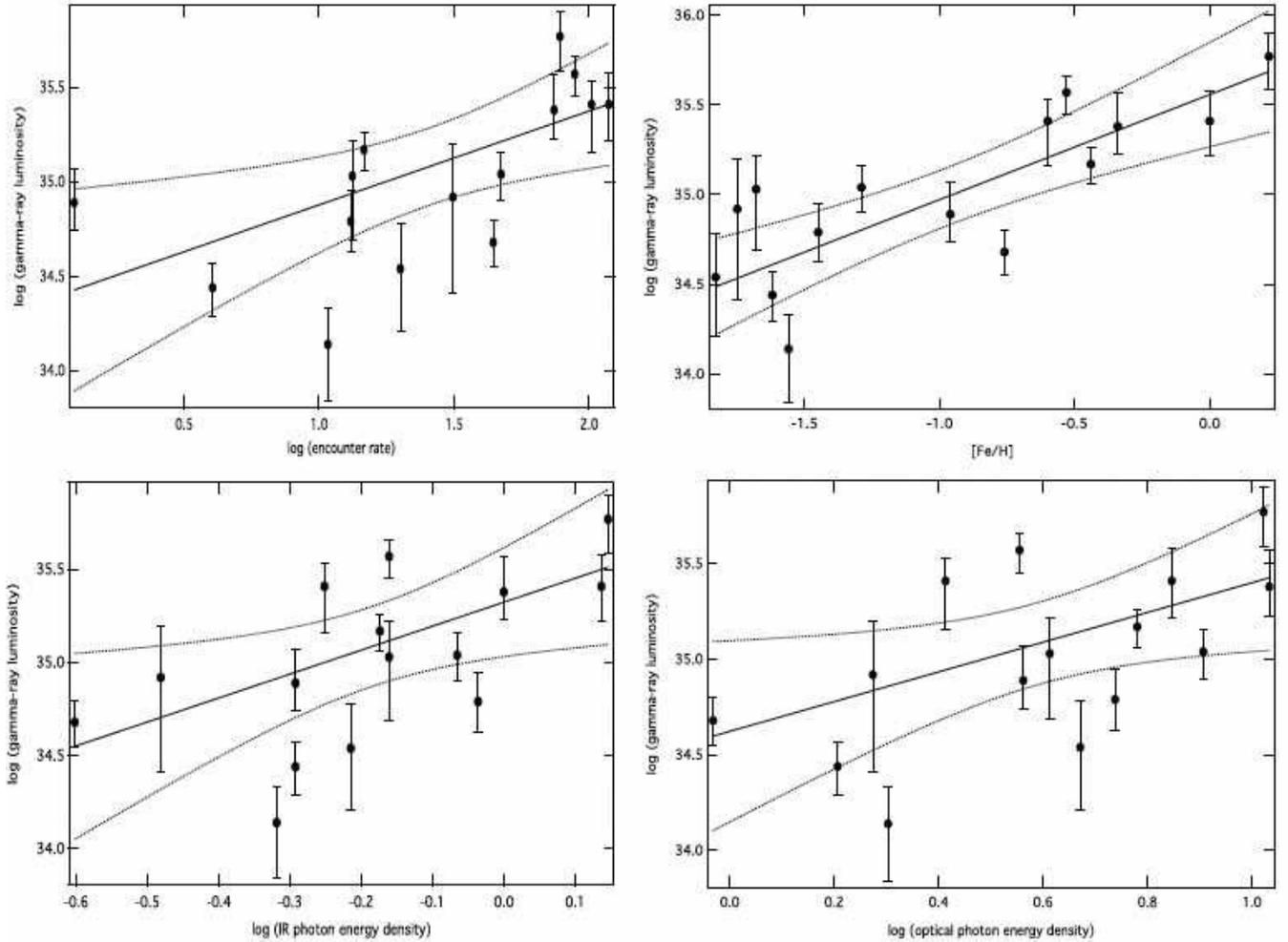,width=18cm,clip=}
\caption[]{$L_{\gamma}$ vs. various individual cluster properties.
The straight lines in the plots represent the best-fits from
the linear regression with the errors of the data points fully taken into account. The dotted lines represent 
the upper and the lower $95\%$ confidence bands.}
\label{correl1}
\end{figure*}

Without a priori knowledge of the distributions of the tested quantities, we follow Pooley et al. (2003) to 
adopt a nonparametric correlation analysis. 
The computed Spearman rank correlation coefficients between $L_{\gamma}$ and various
tested quantities are tabulated in Table~\ref{correl}. 
We have also computed the linear correlation coefficients 
(i.e. Pearson's $r$) for an intutitive account for the data scattering, though they are less robust than 
the Spearman ranks in quantifying the correlations. We have also performed the 1-dimensional regression analysis. 
The best-fit parameters are also given in Table~\ref{correl}. All the quoted uncertainties are 
$95\%$ confidence intervals. 
The best-fit relations of these quantities with $L_{\gamma}$ are plotted as solid straight lines in Figure~\ref{mass}b and 
Figure~\ref{correl1}. We have also plotted the upper and lower $95\%$ confidence bands 
for a visual comparison for the data scattering in each panel. 

Among all these parameters, the weakest correlation is found for the $\log L_{\gamma}-M_{V}$ relation
which is only significant at $\sim15\%$ confidence level. Therefore, there is no convincing correlation 
between these two quantities (see Figure~\ref{mass}b). 
On the other hand, the correlations of $L_{\gamma}$ with $\Gamma_{c}$ and
$\left[{\rm Fe/H}\right]$ are confident at a level over $99\%$ (see Figure~\ref{correl}). All these findings are 
fully consistent with the results from analysing the radio MSP population (Hui et al. 2010).

While the correlation between $L_{\gamma}$ and $\Gamma_{c}$ was reported by Abdo et al. (2010a) with 8 GCs, the 
effect of metalicity was ignored in their work. 
From the $95\%$ confidence bands shown in Figure~\ref{correl1}, the degree of data scattering of the 
$\log L_{\gamma}-\left[{\rm Fe/H}\right]$ relations is found to be the smallest among all 
the tested single parameters. This can be also reflected by the fact that 
its corresponding linear correlation attains the confidence levels over $99.9\%$. 

For the tested soft photon fields, Figure~\ref{correl1} shows that $L_{\gamma}$ also tends to increase with
their energy densities. For $u_{\rm optical}$ and $u_{\rm IR}$, the confidence levels for the
correlations are $>96\%$ and $>99\%$ respectively. 

In comparison to the $\log L_{\gamma}-\log\Gamma_{c}$ and $\log L_{\gamma}-\left[{\rm Fe/H}\right]$ relations,
the relatively large scattering of the data points in the plots
of $\log L_{\gamma}-\log u_{\rm optical}$ and $\log L_{\gamma}-\log u_{\rm IR}$ can be due to the fact that 
the GALPROP code is a simplification of the real situation in the Galaxy, which assumes an axisymmetric distribution
of all parameters of the program (cf. Strong \& Moskalenko 1998). Although it provides reliable average
density of background photons in the Galaxy, the real photon densities at the locations of the GCs can
differ by a factor of a few, in particular for those close to the disk.

Despite the scattering, the correlation analysis strongly suggests $L_{\gamma}$ is likely related to the soft photon 
energy density estimates. This inference is consistent with the ICS modeling the $\gamma-$ray spectra of GCs 
(Cheng et al. 2010) which indicates neither the number of MSPs nor the soft photon energy density is the 
sole factor in determining $L_{\gamma}$. With this consideration, we investigate if $L_{\gamma}$, 
$\Gamma_{c}$/$\left[{\rm Fe/H}\right]$, and $u_{\rm optical}$/$u_{\rm IR}$ span a fundamental plane by a 
2-dimensional regression analysis. We have examined the sample with the following relations:  

\begin{equation}
\log L_{\gamma}=a_{1}+a_{2}\log\Gamma_{c}\\
+a_{3}\log u_{\rm optical}
\end{equation}

\begin{equation}
\log L_{\gamma}=a_{4}+a_{5}\log\Gamma_{c}\\
+a_{6}\log u_{\rm IR}
\end{equation}

\begin{equation}
\log L_{\gamma}=a_{7}+a_{8}\left[{\rm Fe/H}\right]\\
+a_{9}\log u_{\rm optical}
\end{equation}

\begin{equation}
\log L_{\gamma}=a_{10}+a_{11}\left[{\rm Fe/H}\right]\\
+a_{12}\log u_{\rm IR}.
\end{equation}

\begin{figure*}
\centering
\psfig{figure=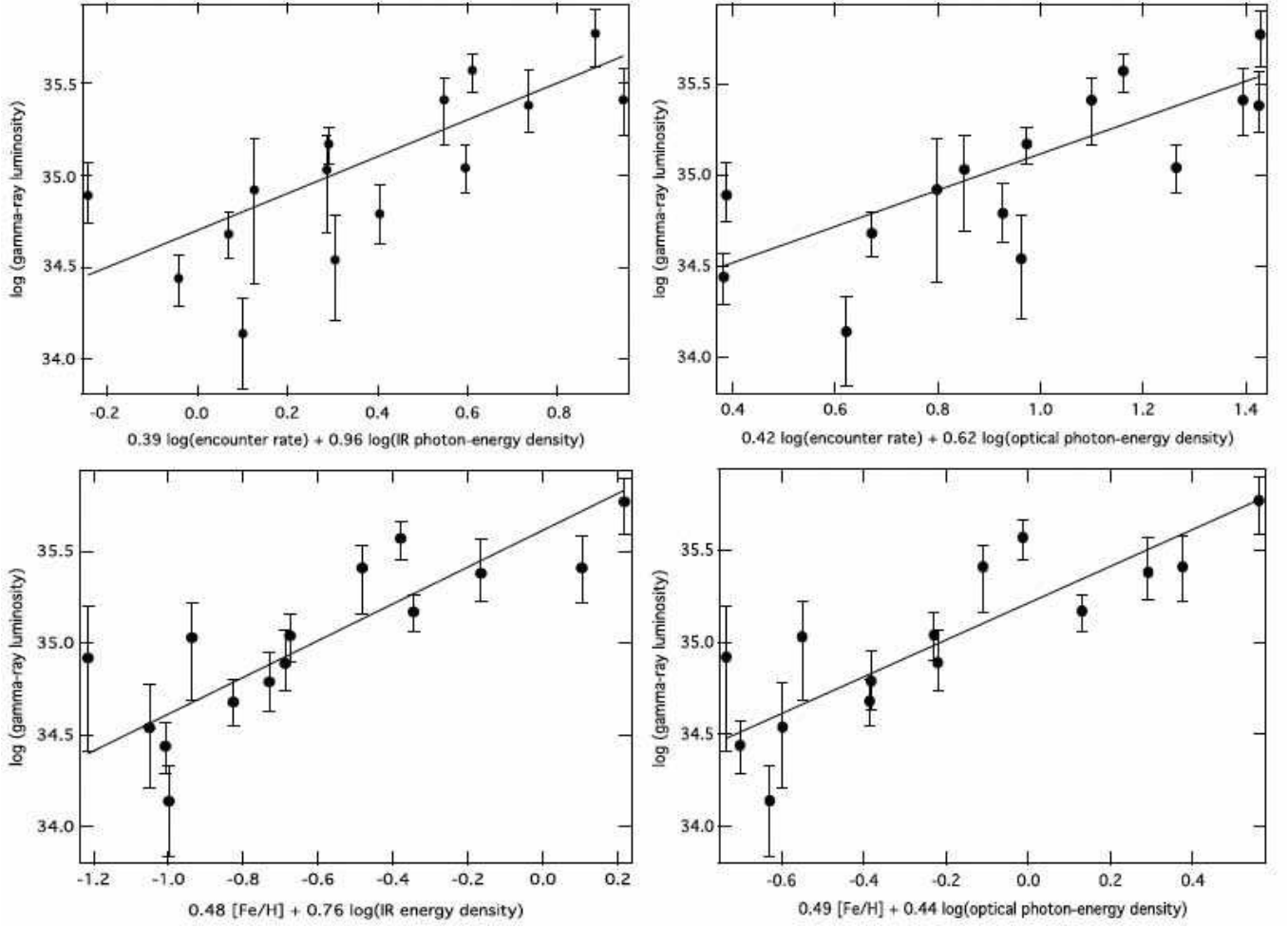,width=18cm,clip=}
\caption[]{The edge-on views of the fundamental plane relations of $\gamma-$ray GCs. 
The straight lines in the plots represent the projected best-fits given in Table~\ref{fp}. }
\label{correl2}
\end{figure*}

\begin{figure*}
\centering
\psfig{figure=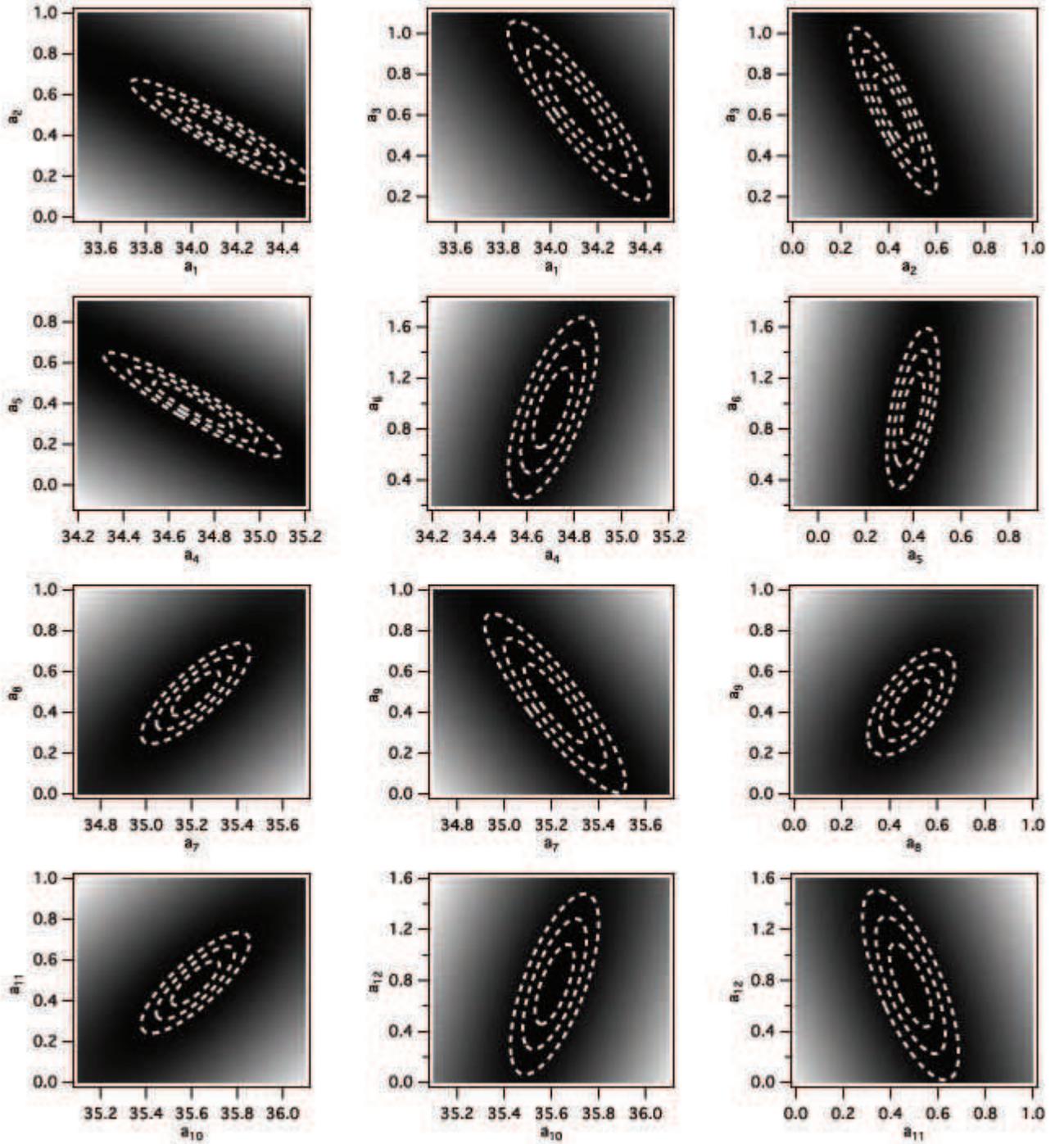,width=18cm,clip=}
\caption[]{The $\chi^{2}$ maps for various parametric spaces of the fundamental plane relations.  
The dashed lines illustrate the $1\sigma$, $2\sigma$ and $3\sigma$ confidence contours for two parameters of interest which 
encircle the best-fit values (i.e. the positions with the lowest $\chi^{2}$).}
\label{contour}
\end{figure*}

The best-fit parameters are tabulated in Table~\ref{fp}. We have shown 
the edge-on view of these best-fit fundamental plane relations in Figure~\ref{correl2}. In comparison with
Figure~\ref{correl1}, the data scatter in these plots are somewhat reduced which suggest these fundmental plane
relations can possibly provide us with better $\gamma-$ray luminosity predictors than the single parameter relations.
To better constrain the uncertainties of these parameters, we have further computed the $1\sigma$, $2\sigma$ and $3\sigma$ 
confidence contours for various parametric spaces which are shown in Figure~\ref{contour}. 

\begin{center}
\begin{deluxetable}{c|cc|cc|cc}
\tablewidth{0pc}
\tablecaption{Correlation and 1-D regression analysis of $\log L_{\gamma}$ versus various
cluster properties.}
\startdata
\hline\hline
Parameters & Spearman rank & Prob$_{\rm S}$\tablenotemark{a} & Pearson's $r$ & Prob$_{\rm P}$\tablenotemark{b} & $m$\tablenotemark{c}
& $c$\tablenotemark{c}\\\hline\hline
$\log\Gamma_{\rm c}$         & 0.7918 & 0.9996 & 0.6414 & 0.9900 & $0.50\pm0.16$ & $34.38\pm0.25$ \\
$\left[{\rm Fe/H}\right]$    & 0.7614 & 0.9990 & 0.7912 & 0.9996 & $0.59\pm0.15$ & $35.56\pm0.15$ \\
$M_{V}$                      & -0.0536 & 0.1505 & -0.0767 & 0.2141 & $0.04\pm0.08$ & $35.40\pm0.67$ \\
$\log u_{\rm optical}$       & 0.5523 & 0.9672 & 0.5976 & 0.9814 & $0.78\pm0.27$ & $34.62\pm0.19$\\
$\log u_{\rm IR}$            & 0.6589 & 0.9925 & 0.5970 & 0.9812 & $1.29\pm0.44$ & $35.33\pm0.12$
\enddata
\tablenotetext{a}{The probability that the Spearman rank correlation coefficient is different from zero.}
\tablenotetext{b}{The probability that the linear correlation coefficient (i.e. Pearson's $r$) is different from zero.}
\tablenotetext{c}{The best-fits for $\log L_{\gamma}=mx+c$ where $x$ is the corresponding parameters listed in column 1.}
\label{correl}
\end{deluxetable}
\end{center}

\begin{center}
\begin{deluxetable}{cc}
\tablewidth{0pc}
\tablecaption{Best-fit fundamental plane relations of $\gamma-$ray GCs.}
\startdata
\hline\hline
Parameters & Best-fit values\\\hline\hline
$a_{1}$ & $34.12\pm0.29$ \\
$a_{2}$ & $0.42\pm0.17$ \\
$a_{3}$ & $0.62\pm0.29$ \\\hline
$a_{4}$ & $34.70\pm0.30$ \\
$a_{5}$ & $0.39\pm0.18$ \\
$a_{6}$ & $0.96\pm0.49$ \\\hline
$a_{7}$ & $35.21\pm0.29$ \\
$a_{8}$ & $0.49\pm0.18$ \\
$a_{9}$ & $0.44\pm0.31$ \\\hline
$a_{10}$ & $35.61\pm0.16$ \\
$a_{11}$ & $0.48\pm0.17$ \\
$a_{12}$ & $0.76\pm0.50$
\enddata
\label{fp}
\end{deluxetable}
\end{center}

\section{DISCUSSION}

We have examined the $\gamma-$ray emission properties of a group of GCs. By investigating the possible
correlations between the $\gamma-$ray power and a number of cluster properties, we shed light on the
origin of the $\gamma-$rays from these GCs. First of all, the correlation between $L_{\gamma}$ and
$\Gamma_{c}$ suggests the high energy radiation are intimately related to
the population of dynamically-formed objects, which are presumably MSPs, confirming Abdo et al. (2010a) who 
used 8 GCs in their study. Together with the lack of any
correlation with $M_{V}$ and hence the cluster mass, this is fully consistent with the inference suggested by 
Hui et al. (2010) and consolidates the dynamical formation scenario of MSPs in GCs. 

Apart from $\Gamma_{c}$, we have found that $L_{\gamma}$ is also positively correlated with
$\left[{\rm Fe/H}\right]$. This is well-consistent with the tendency deduced from studying the radio MSP
population in GCs (Hui et al. 2010) and the fact that the GC possesses the highest [Fe/H] also has the highest 
$L_{\gamma}$ (Tam et al. 2010). Ivanova (2006) proposes that the absence of the outer
convective zone in metal-poor main sequence donor stars in the mass range of $0.85\msun$ - $1.25 \msun$,
in comparison to their metal rich counterparts can be responsible, since the absence of magnetic
braking in such stars precludes orbital shrinkage, thereby, significantly reducing the binary parameter
space for the production of bright LMXBs. For a conventional scenario, MSPs are the old pulsars that
have passed through the death-line in $P-\dot{P}$ diagram which are subsequently spun-up in the binaries.
As the metalicity determines the parameter space for successful Roche-lobe overflow, it is also a key parameter
in determining the intrinsic number of MSPs in a GC (Hui et al. 2010; Ivanova 2006).

We note that the link between the LMXBs in extragalactic GCs and the metalicity is somewhat weaker than with 
the cluster mass (e.g. Sivakoff et al. 2007; Kim et al. 2006; Kundu et al. 2002), which is different from the inference
drawn from our investigation of the Galactic MSP-hosting or $\gamma-$ray selected clusters. However, a direct comparison 
between these two populations has to be cautious. As the MSPs are long-lived and are produced by the previous generations 
of LMXBs, their dynamical properties might be different from that of the LMXB population currently observed. 
Since the relaxation time at the cluster core is generally longer than 
the lifetime of LMXBs, the cluster is continuously evolved with mass segregation at the cluster center which 
can result in a varying formation rate of compact binaries (cf. the discussion in Hui et al. 2010). 
Also, while a large number of LMXB-hosting GCs in Virgo cluster early-type galaxies have relaxation times 
$>2.5$~Gyr, there is no single GC in our Galaxy with a relaxation timescale larger than this value contains an active LMXB 
(cf. Sivakoff et al. 2007). Although the reason is still unclear, this suggests possible different properties between 
the Milky Way GCs and the extragalactic ones. Further investigations are required to understand the difference.

It is instructive to compare the fundamental plane relations of the $\gamma-$ray population with the 
best-fits inferred from the radio MSP population in GCs. Hui et al. (2010) have found that the 
slopes of $\log\Gamma_{c}$ and $\left[{\rm Fe/H}\right]$ inferred from the radio GC MSPs population are $0.69\pm0.11$ and 
$0.72\pm0.11$ respectively. Within these quoted uncertainties, the slope of the $\left[{\rm Fe/H}\right]$ 
relation for the radio population is found to intersect with the $2\sigma$ error contours for the corresponding 
parameters of the $\gamma-$ray fundamental plane relations (i.e. $a_{8}$ and $a_{11}$). On the other hand, the logarithmic 
slope of the $\Gamma_{c}$ relation for the radio population is only marginally overlapped with the rims of the $3\sigma$ 
error contours for the corresponding parameters inferred from the $\gamma-$ray population (i.e. $a_{2}$ and $a_{5}$). 

We have also identified possible positive correlations with various soft photon fields which have 
significances compatible with those for the encounter rate and the metalicity. These correlations 
are not expected from the magnetospheric model. Together with the uncertainty of the sustainability of 
the outergaps in the MSPs in GCs (see \S1), our finding further motivate the exploration of alternative explanations  
for the origin of the observed $\gamma$-ray from GCs. 

Abdo et al. (2010a) have argued that the $\gamma$-ray emission is magnetospheric in nature because of the 
hard photon indices and the cutoff energies inferred from the phenomenological model is consistent 
with the values expected from the magnetospheric model. On the other hand, Cheng et al. (2010) have recently 
found that the ICS model can also describe the observed $\gamma-$ray spectra of all the GCs discovered by Abdo et al. 
(2010a) very well. Simply based on the model fitting, we were not able to discriminate these two scenarios 
unambiguously. However, different from the case of the magnetospheric model,
positive correlation between the energy density of the soft photon fields are expected in a ICS scenario as
the ICS power is directly proportional to soft photon energy density. 

The energy density of the background soft photon field depends on the location of the cluster. We notice 
these $\gamma$-ray GCs are possibly resided in the Galactic bulge and therefore they are also metal-rich clusters. 
This results in a natural correlation between the metalicity and the soft photon energy density with a significance $>95\%$. 
Therefore, it is non-trivial to disentangle the effects of these two parameters.  

In any case, our investigation strongly suggests that either the metalicity or 
the soft photon energy density has to be the new parameter, 
in addition to $\Gamma_{c}$, in determining the observed $\gamma$-ray luminosities.  
This inference is supported by comparing the results reported 
by Abdo et al. (2010a) and Tam et al. (2010). Apart from the 8 confirmed cases, Abdo et al. (2010a) have also reported 
5 non-detections which include three upper-limits and two other cases with the $\gamma-$ray emission slightly offset 
from the respective GC cores. 
With the LAT data of a longer exposure, Tam et al. (2010) have found a larger number of $\gamma$-ray GCs 
including 4 previously non-detected cases in Abdo et al. (2010a). 
This leaves M~15 to be the only non-detected GC in Abdo et al. (2010a). This 
is not unexpected from trends of metalicity and background soft photon energy density. Although the encounter rate of 
M~15 ($\Gamma_{c}=53.9$) is even higher than M~62, its metalicity ([Fe/H]=-2.26) and the background soft energy densities 
at its location ($u_{\rm optical}=0.44$~eV~cm$^{-3}$; $u_{\rm IR}=0.11$~eV~cm$^{-3}$) are much lower than those of 
all 15 confirmed $\gamma$-ray GCs. 

In view of the aforementioned complication, it is not possible to discriminate the ICS and the 
magnetospheric scenarios unambiguously simply based on the currently available information. 
Also, there is still a degeneracy within the context of ICS model. 
Cheng et al. (2010) show that the $\gamma-$ray spectrum from 47~Tuc can be explained equally well by
upward scattering of either the relic photons, the Galactic infrared photons or the Galactic optical
photons whereas the $\gamma-$ray spectra from the other seven GCs reported by Abdo et al. (2010a) are 
best fitted by the upward scattering of either the Galactic infrared photons or the Galactic optical photons. 
This has prompted us also to discriminate which source provides the predominant soft photon field for ICS. 

Since the IC radiation power is directly proportional to the energy density of the soft photon field, a 
logarithmic slope of unity is thus expected for the fundamental plane parameters $a_{3}$, $a_{6}$, $a_{9}$ and 
$a_{12}$. For the corresponding parameters of $u_{\rm IR}$ (i.e. $a_{6}$ and $a_{12}$), the line of unity is found to 
cut through the centers of the $1\sigma$ error contours for both $\Gamma_{c}$ and [Fe/H] fundamental plane relations. 
On the other hand, for the parameters correspond to $u_{\rm optical}$ (i.e. $a_{3}$ and $a_{9}$), the expected value 
is only marginally intersected with their $3\sigma$ error contours. Although an unambiguous conclusion cannot be drawn 
from the current population yet, the comparison between the theoretical expectation and the fundamental plane parameters 
does favor the scenario involving the background infrared emission as the soft photon field for IC upscattering. 

We would like to point out that the predicted spectral shape in energy regime much larger than 10 GeV 
is significantly different for different soft photon fields (cf. Fig.~2 in Cheng et al. 2010). 
Therefore, observations with TeV facilities, such as MAGIC, HESS and VERITAS, can be feasible to lift up this degeneracy.
Furthermore, as the ICS model and the magnetospheric model predict a rather different TeV spectrum for GCs 
(Cheng et al. 2010; Venter et al. 2009), TeV observations in the future can possibly better 
discriminate these two possible contributions of soft photons. 

Constraints for the emission model can also be derived from the other energy bands. 
With the diffusion of the relativistic pulsar wind particles, it has been shown that extended 
radio and X-ray emission from the GCs can also be produced by synchrotron radiation and ICS respectively 
(Cheng et al. 2010). This is consistent with the recent discovery of the diffuse X-rays around Terzan~5 which exteneded up 
$\sim10$~pc (Eger, Domainko \& Clapson 2010). Although a clear scenario cannot be identified yet, these diffuse X-rays are 
more likely to have a non-thermal origin (Eger, Domainko \& Clapson 2010). Assuming these X-rays are originated from the 
tail of ICS, the corresponding $\gamma-$ray spectrum can be calculated (Cheng et al. 2010). Therefore, a systematic search 
for the extended X-ray and radio feature outside the half-mass radii of the other $\gamma-$ray GCs can provide us indpendent 
constraints.

In exploring the fundamental plane relations, our analysis suggests that by combining the soft photon energy densities with
$\Gamma_{c}$/$\left[{\rm Fe/H}\right]$ the data scattering can be reduced.
These best-fit relations can provide the indicators in 
identifying what kind of GCs are potential $\gamma-$ray sources for a further search. And the other way round,
any deeper $\gamma-$ray search from the GCs can result in an enlarged sample size and a lower sensitivity limit than 
the current value (i.e. $6\times10^{-12}$~erg~cm$^{-2}$~s$^{-1}$), which will certainly enable a further test for all 
these reported relations.

\acknowledgments{
The authors would like to thank Kinwah Wu and anonymous referee for the useful 
discussion and providing comments for improving the quality of this manuscript.
CYH is supported by research fund of Chungnam National
University in 2010.
KSC is supported by a GRF grant of Hong Kong Government
under HKU700908P. DOC and VAD are supported by
the RFBR grant 08-02-00170-a,  
the NSC-RFBR Joint Research Project RP09N04 and
09-02-92000-HHC-a. 
And AKHK is supported partly by the National
Science Council of the Republic of China (Taiwan)
through grant NSC96-2112-M007-037-MY3 and a Kenda
Foundation Golden Jade Fellowship.}

\end{document}